\documentclass[%
 aip,
 amsmath,amssymb,
 reprint,%
]{revtex4-1}

\usepackage{graphicx}% Include figure files
\usepackage{dcolumn}% Align table columns on decimal point
\usepackage{bm}% bold math
\usepackage[utf8]{inputenc}
\usepackage[T1]{fontenc}
\usepackage{mathptmx}
\usepackage{etoolbox}
\usepackage{float}
\usepackage{siunitx}
\usepackage{tikz}
\usepackage{siunitx}
\usepackage{booktabs}
\usepackage{tabularx}
\usepackage{lipsum}
\usepackage{array}
\usepackage{makecell}
\usepackage{bigstrut}

\makeatletter
\def\@email#1#2{%
 \endgroup
 \patchcmd{\titleblock@produce}
  {\frontmatter@RRAPformat}
  {\frontmatter@RRAPformat{\produce@RRAP{*#1\href{mailto:#2}{#2}}}\frontmatter@RRAPformat}
  {}{}
}%
\makeatother
\begin{document}

\preprint{AIP/123-QED}

\title[Realization of a chip-based hybrid trapping setup for $^{87}$Rb atoms and Yb$^{+}$ Ion crystals]{Realization of a chip-based hybrid trapping setup for $^{87}$Rb atoms and Yb$^{+}$ Ion crystals}
% Force line breaks with \\
 % \altaffiliation[Current affiliation: ]{Department of Electrical and Computer Engineering, University of California, Los Angeles, Los Angeles, CA 90095, USA}
 \author{A. Bahrami}
\email{abasalt@ucla.edu.}
\affiliation{ 
Department of Electrical and Computer Engineering, University of California, Los Angeles, Los Angeles, CA 90095, USA%\\This line break forced with \textbackslash\textbackslash
}%
\affiliation{%
Institut f{\"u}r Physik, Johannes Gutenberg-Universit{\"a}t Mainz, Mainz D-55099, Germany%\\This line break forced% with \\
}%

% \author{M. M{\"u}ller}
% \affiliation{%
% Institut f{\"u}r Physik, Johannes Gutenberg-Universit{\"a}t Mainz, Mainz D-55099, Germany%\\This line break forced% with \\
% }%

\author{F. Schmidt-Kaler}
 % \homepage{http://www.Second.institution.edu/~Charlie.Author.}
\affiliation{%
Institut f{\"u}r Physik, Johannes Gutenberg-Universit{\"a}t Mainz, Mainz D-55099, Germany%\\This line break forced% with \\
}%

\date{\today}% It is always \today, today,
             %  but any date may be explicitly specified

%   ABSTRACT MUST NOT BE MORE THAN 250 WORDS
\begin{abstract} 
Hybrid quantum systems integrate laser-cooled trapped ions and ultracold quantum gases within a single experimental configuration, offering vast potential for applications in quantum chemistry, polaron physics, quantum information processing, and quantum simulations. In this study, we introduce the development and experimental validation of an ion trap chip that incorporates a flat atomic chip trap directly beneath it. This innovative design addresses specific challenges associated with hybrid atom-ion traps by providing precisely aligned and stable components, facilitating independent adjustments of the depth of the atomic trapping potential and the positioning of trapped ions. Our findings include successful loading of the ion trap with linear Yb$^{+}$ ion crystals and the loading of neutral $^{87}$Rb atoms into a mirror magneto-optical trap (mMOT).
\end{abstract}

\maketitle

% \begin{quotation}
% The ``lead paragraph'' is encapsulated with the \LaTeX\ 
% \verb+quotation+ environment and is formatted as a single paragraph before the first section heading. 
% (The \verb+quotation+ environment reverts to its usual meaning after the first sectioning command.) 
% Note that numbered references are allowed in the lead paragraph.
% %
% The lead paragraph will only be found in an article being prepared for the journal \textit{Chaos}.
% \end{quotation}

\section{\label{sec:Introduction}Introduction}
To advance in the field of hybrid quantum systems, we employ a combination of trapping methods for atomic ions and neutral atoms. The trapped ions provide highly controllable quantum systems, making them a valuable platform for a wide range of applications, including quantum information~\cite{Haffner.2008, Singer.2010}, high-resolution spectroscopy, and tests of fundamental physics~\cite{Safronova:2018}. On the other hand, interactions between neutral atoms primarily result from short-range van der Waals forces, which range from one to several angstroms.

Atomic ions are confined in Paul traps~\cite{Paul:1958,Raizen:1992}. These traps employ a radio frequency (RF) quadrupolar field to create a harmonic trapping potential. In linear Paul traps, this quadrupolar field is extended along a line, and additional electrodes are used to precisely control of the ion crystal position and shape. Neutral atoms have been confined using various techniques such as magneto-optical traps (MOT)\cite{Migdall:1985}, magnetic quadrupolar fields\cite{Pritchard.1983}, or far-detuned laser light~\cite{Kuppens.2000}. We are combining different trapping methods to build the device for a hybrid atom-ion system to exploit its unique properties. The temperature of a confined atomic cloud can be reduced by combinations of laser and evaporation cooling to nanokelvin, even pico Kelvin range, while the temperature of Doppler cooled trapped ions is about a milli Kelvin, advanced cooling techniques may reach a few tenths of micro Kelvin. Sympathetic cooling an ion crystal by means of thermalization with atoms was one of the early motivations for such hybrid systems~\cite{Krych:2013,Zipkes:2010b, Meir:2016}. However, the RF-driven motion of a trapped ion during the interaction with the neutral atom, a short-range (Langevin) collision, leads to heating. Research has shown that the lowest temperatures can be reached for the largest ion-atom mass ratios $m_{i}/m_{a}$\cite{Cetina:2012}. The Yb$^+$/Li hybrid system with a mass ratio of $m_{i}/m_{a}\approx 29$ may be considered optimum to minimize that adverse effect. For other combinations of atoms and ions, such as Rb/Ba$^+$, Rb/Rb$^+$, Rb/Yb$^+$, Rb/Sr$^+$, ect. where short-range collision-induced heating has prohibited to enter low energy scattering regimes. 

We present a chip-based hybrid trapping device as a novel tool to largely avoid collision-induced heating for any combination of atom and ion. Two properties are key: first, the device integrates a linear Paul trap where ions crystals may be positioned accurately with respect to a Ioffe-Pritchard (IP) magnetic trap that confines an atomic sample. Second, chip-based atom or ion traps feature tightly confining potentials such that both, such that the ion crystal may be positioned {\it close}, but not directly on the atomic sample. In such way, one may explore interactions at mutual distances of a $\mu$m, but short-range collision-induced heating is mitigated. 

Here, we describe a device for a combination of Rb atoms and Yb$^+$ ions \cite{Abasalt:2019}. After giving an overview of the entire device, we are describing the ion chip, then the atom chip trap, followed by a short description of laser systems to cool the ions and atoms, and optics to detect their laser induced fluorescence. The planar Paul trap chip is sitting directly on top of the atom chip trap device. We show the operation of the planar segmented ion trap with a linear ion crystal \cite{chiaverini.2005}. Also we demonstrate the operation of a mMOT~\cite{Reichel:1999}, to load the atomic chip trap which features u- and z-shaped wires for versatile magnetic trapping situations.

% SYSTEM OVERVIEW %%%%%%%%%%%%%%%%%%%%%%%%%%%
\section{System Overview}
% fsk: the overview of the entire device is missing, add both as sect I. and add the overview scheme

Our experimental setup is a hybrid atom-ion chip trap designed to capture and manipulate both laser-cooled neutral $^{87}$Rb atoms and Yb$^+$ ions. This system allows for controlled investigations into the interactions of these particles in a precisely defined environment.

		\begin{figure*}
		\centering
		\includegraphics[width=6.69in]{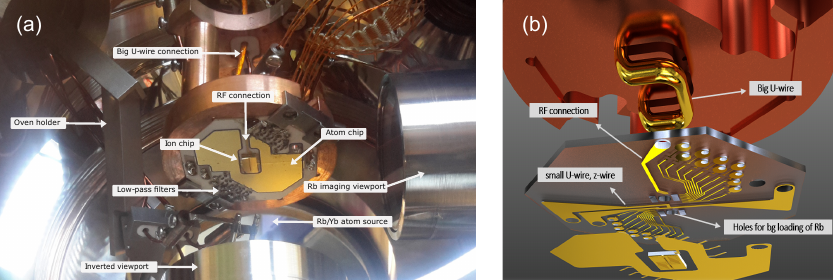}
		\caption{A visual representation of our experimental setup when installed within an ultra-high vacuum (UHV) chamber (a). The overall dimensions of the chip are 45\,mm in outer diameter and 1\,mm in total height. As the trapping area is limited by the surface of the ion trap at a position of 0.6\,mm, the trap depth is estimated to be 273\,$\mu$K. (b) presents an exploded view of the atom-ion chip assembly, incorporating small u- and z-shaped wires for transferring and confining the mMOT in the final stage. The z-shaped wire has a height of 0.08\,mm and width of 0.6\,mm. }
		\label{fig:CompleteTrap}
		\end{figure*}

\begin{figure*}[ht]
\centering
\includegraphics[width=6in]{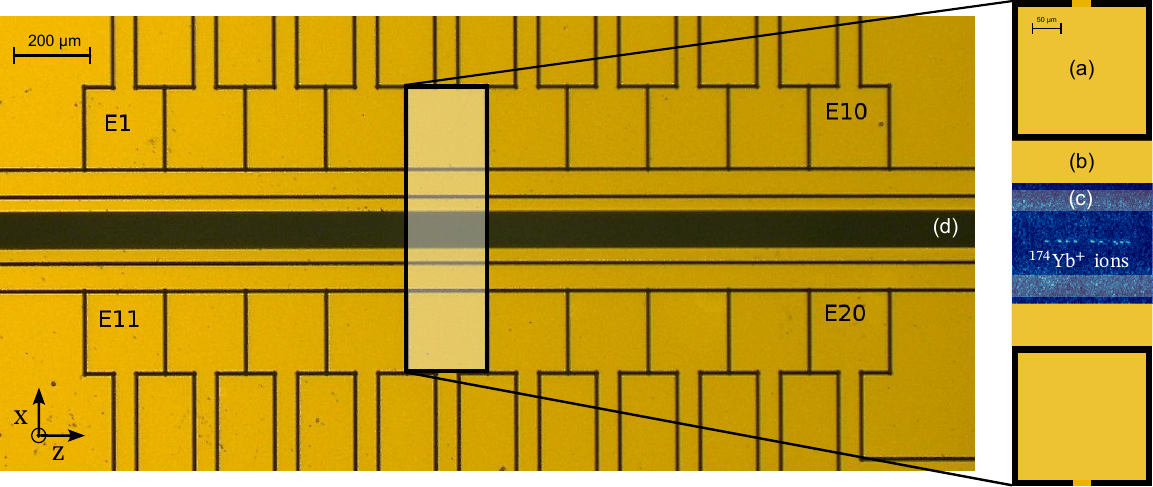}
\caption{The image above is a magnified optical microscope image of the microfabricated surface trap used in our experimental setup. It features 21 static voltage electrodes, including 20 with a size of 200$\times$200\,$\mu$m$^{2}$ (a), which are used for radial confinement (E01 - E20), one long, symmetric F-shaped rail for RF confinement (b) and one inner compensation electrode (c), which extends axially and symmetrically along a slit of 100\,$\mu$m width and 5\,mm length (d). This slit is employed to load Rb atoms from the dispenser positioned directly behind the ion trap. In the experiments described here, Yb$^{+}$ ion crystals are trapped and confined along the trap axis (z-direction). The CCD image, captured with an exposure time of 1.3\,s, depicts nine $^{174}$Yb$^{+}$ ions in a linear crystal that is trapped with corresponding trap frequencies of ($\omega_{x},\omega_{z})/2\pi = (406, 110)$\,kHz.}
\label{fig:iontrap}
\end{figure*}

The experimental setup comprises an atom chip designed for trapping neutral $^{87}$Rb atoms. This chip incorporates an mMOT for loading and cooling, current-carrying wires for atom transport, and a magnetic trap (MT) for the final confinement of atoms (Fig. \ref{fig:CompleteTrap}). Additionally, a surface electrode ion trap (Fig.~\ref{fig:iontrap}) is employed for trapping Yb$^+$ ions. Underneath the atom chip we have implemented a big u-shaped wire essential for capturing atoms in a secondary mMOT (Fig.~\ref{fig:CompleteTrap}b). The atom chip also features small wires with two distinct geometries: a u-shaped configuration designed to compress and transfer atoms from the second mMOT to the ion trapping region, and a z-shaped wire designed for trapping atoms in MT. The atom chip itself is fabricated using thick film technology, enabling the printing of UHV compatible, sub-millimeter scale electrical circuits on an alumina substrate (Al$_{2}$O$_{3}$).
Additionally, the atom chip incorporates a filter board for the direct current control electrodes of the ion trap, which incorporates 3.38\,MHz low-pass filters (with capacitance of 4.7\,nF and resistance of 10\,$\Omega$) for all DC electrodes to mitigate RF pickups in proximity to the ion trap.

\begin{figure}[H]
\centering
\includegraphics[scale=1]{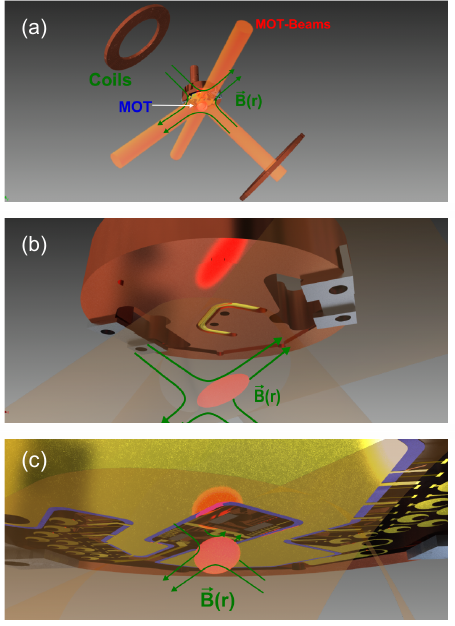}
\caption{(a) An mMOT is created using $\sigma^{\pm}$ light beams and a magnetic quadrupole field generated by two coils carrying current in an anti-Helmholtz configuration. (b) The quadrupole field for the mMOT is produced by the magnetic field of a big u-shaped wire carrying current beneath the chip, combined with a uniform bias field. (c) On the chip, small u-shaped wire forms a more compact and steeper quadrupole field directly above the ion trap.}
\label{fig:mMOT_stages}
\end{figure}

After capturing atoms in the mMOT (Fig.~\ref{fig:mMOT_stages}a), the mMOT coils switch to a bias field while activating the large u-shaped wire beneath the atom chip (Fig.~\ref{fig:mMOT_stages}b). As a result, the atoms can be confined approximately 2-5\,mm below the ion chip surface within a smaller spatial area.

Following this, the atoms are transferred to a potential created by the small u-shaped wire on the atom chip (Fig.~\ref{fig:mMOT_stages}c). During this process, we move the atoms close to the ion trap surface, compressing the atom cloud into a smaller and steeper mMOT volume \cite{petrich1994behavior}. In the final step, an IP trap is established by turning on the z-shaped wire (Fig.~\ref{fig:z_u_wires}) on the atom chip along with a bias magnetic fields. $^{174}$Yb$^+$ ions are loaded into the ion trap, creating a shared volume where the interactions between the atoms and ions can be studied in detail. To maintain the required low pressure, we employ an ion getter pump, ensuring a stable experimental environment with pressures as low as $1.3 \times 10^{-9}$\,mBar.

The setup should allow us to trap \(^{87}\text{Rb}\) atoms which will be evaporated from an alkali metal dispensers with temperatures of about \(750 \,^{\circ}\text{C}\) and to cool them down to \(T_D = 140 \, \mu\text{K}\). The final trap frequencies for trapped atoms are $\omega/2\pi \sim$ (1.17, 1.17, 0.08) \,kHz.

\section{ion trap description}
% fsk: sect. II will be the ion trap description, and fig 1.

This ion trap is a microfabricated surface electrode segmented design with a trapping height of 100\,$\mu$m. The dimensions of the chip are 9\,mm in length, 4.5\,mm in width, and 500\,$\mu$m in thickness. This chip includes a slit of a 100\,$\mu$m wide and 6.5\,mm long, positioned just below the ion trapping region. This slit will function as the entry point for loading atoms using an atomic dispenser.

The fabrication process of this chip follows these steps: The electrode structure is etched onto a fused silica substrate using a combination of laser weakening and HF etching. Subsequently, four metal layers are evaporated onto the substrate. These metal layers consist of 20\,nm of Titanium, 150\,nm of gold, another 20\,nm of Titanium, and an additional 150\,nm of gold~\cite{hellwig2010fabrication}. The chip structure comprises 21 DC electrodes labeled as E1 to E21, along with one RF electrode (RF), while the rest of the surface remains grounded.  The isolation between the electrodes is approximately 10\,$\mu$m wide and 50\,$\mu$m deep, which is large enough to prevent electrical breakdown at 100-200\,V$_{pp}$. To provide the voltage for the DC electrodes, we use FPGA-controlled digital-to-analog conversion (DAC) boards \cite{walther2012controlling}. It supports up to 48 independent analog channels (16 bits) and 25 digital TTL signals per unit. Analog channels have a voltage range of $-10 \, \text{V}$ to $+10 \, \text{V}$ with a resolution of $0.12 \, \text{mV}$ per least significant bit (LSB).

The ion trap is glued to the center of the atom trap using a UV-curable adhesive. To establish the necessary electrical connections from the filter board to the ion trap, a wire bonder (HB10) is used. The atom chip is coated with a top layer of gold, which serves as a protective layer for the wires as well as a reflecting mirror for the mMOT laser beams. Typically, we trap \( \text{Yb}^+ \) ions at a distance of 100\,$\mu$m from the trap surface, with the ability to position them axially with an accuracy of $\pm$\,80\,nm.

\section{Atom traps}
% sect III will be the neutral atom traps. first an overview of the different kind of traps, mirror MOT and Joffe P. Add pictures of the U and Z wire. The part of the MOT you can take in large parts.

% https://arxiv.org/pdf/1310.6054.pdf
% https://sci-hub.se/https://journals.aps.org/pra/abstract/10.1103/PhysRevA.58.R2664
% https://arxiv.org/pdf/cond-mat/9904034.pdf
Magnetic trapping of atoms involves employing magnetic fields to confine and manipulate atoms. One common technique is the use of a quadrupole magnetic trap, established by employing two coils with currents flowing in opposite directions \cite{esslinger1998bose}. However, quadrupole trap has a zero minimum at
the origin and significant limitation of this traps is the loss of cold atoms from the trap due to nonadiabatic \textit{spin-flip Majorana transition}~\cite{davis1995evaporative}. This problem has been addressed through the implementation of time-averaged orbiting potential (TOP) traps \cite{petrich1995stable} and IP magnetic traps~\cite{ernst1998bose}.  The TOP trap operates by leveraging the time-averaging of an adiabatically modulated trapping potential. In this approach, the potential is varied slowly to ensure adiabatic conditions, allowing for the stable trapping of neutral atoms. Conversely, the IP trap relies on a combination of static and radiofrequency (RF) magnetic fields to create a three-dimensional trapping potential. The RF field is typically employed for radial confinement, while the static field contributes to axial confinement. 

		\begin{figure}[ht]
		\centering
		\includegraphics[width=3.37in]{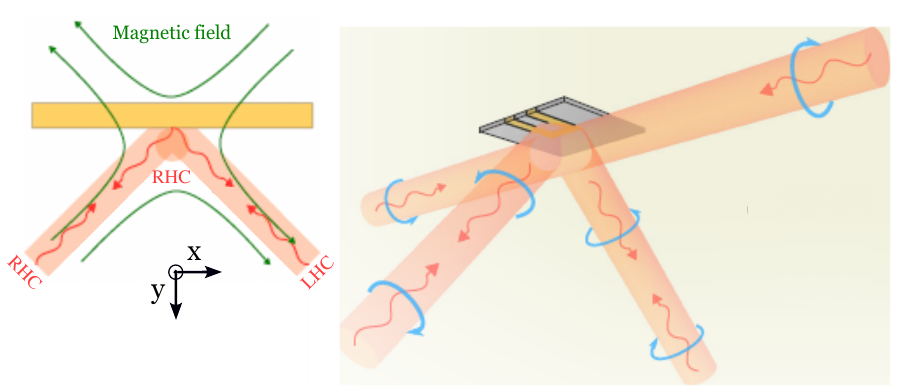}
		\caption{Laser beams (red) with RHC and LHC laser lights to run the $\sigma^{+}$ and $\sigma^{-}$ transitions with a quadrupole magnetic field (green). The gold-coated surface of the atom trap is depicted by the yellow rectangle.}
		\label{fig:QPbeams}
		\end{figure}
  
To create a mMOT, we used right-hand circular (RHC) and left-hand circular (LHC) polarized light to drive the $\sigma^{+}$ and $\sigma^{-}$ transitions (Fig.~\ref{fig:QPbeams}). To trap and cool atoms we used two home-made external cavity diode lasers (ECDLs) that have a minimum output power of 130\,mW under continuous wave conditions (LNC728PS01WW). Each laser is individually frequency-stabilized to the atomic transitions in $^{87}$Rb's D2 line with a natural linewidth of $\Gamma \approx 2\pi \times 6.0$ \, \text{MHz} \cite{volz1996precision}. These lasers provide the necessary cooling, red-detuned from $|\text{F}=2\rangle \rightarrow |\text{F}'=3\rangle$ resonant transition, and repump laser frequencies, $|\text{F}=1\rangle \rightarrow |\text{F}'\rangle$ resonant transition, for the experiment. 

% Time-of-flight images are captured with a probe beam resonant with the $|\text{F}=2\rangle \rightarrow |\text{F}'=3\rangle$ transition.

% The laser configuration incorporates a reflective grating (GH13-18V: Visible Reflective Holographic Grating, 1800/mm, 12.7\,mm x 12.7\,mm x 6\,mm) placed in Littrow configuration ~\cite{hawthorn2001littrow}. Additionally, this laser design comprises a collimation tube and an aspheric lens (C230TMD-A: $f = 4.51$\,mm, $\text{NA} = 0.55$, Mounted Aspheric Lens, ARC: 350\,-\,700\,nm) to collimate the outgoing laser beam (LT110P-B: $f = $6.24\,mm, $\text{NA} = 0.40$, AR Coated: 650\,-\,1050\,nm) and a decoupling mirror (Tafelmeyer float glass HR/11 E). The laser housing is sealed with an aperture window (WG11050-A: N-BK7 Broadband Precision Window, AR Coated: 350 - 700nm, $t = 5$\,mm). A Peltier element (Quick-Cool QC-71-1.4-8.5M) beneath the laser diode mount stabilizes the temperature of the laser diode. To adjust the laser frequency to the desired values, we vary the voltage applied to the piezo behind the reflective grating, which changes the length of the external cavity. Additionally, the temperature and current of the laser diode can be adjusted.

\begin{figure}[h]
	\centering
	\includegraphics[scale=1.1]{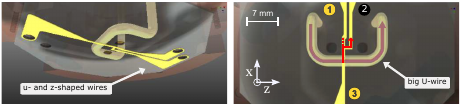}
	\caption{Small u- and z-shaped wires to compress and confine atoms in an IP trap. Grounded at (2), we pass current through (1) to generate u-shaped wire potential and through (3) for the z-shaped wire potential. Red arrows illustrate the flow of electric currents.}
	\label{fig:z_u_wires}
\end{figure}

The atom trap chip features two primary trapping wires: a small u- and a z-shaped (Fig.~\ref{fig:z_u_wires}). During the final trapping stage with the z-shaped wire, for a field zero-minimum at $(x = 0,\, y = 0.7,\, z = 0)$\,mm, we applied bias fields of $\textit{\textbf{B}}_{\text{bias}} = -\textit{\textbf{B}}_{\text{z-wire}}(0, \,0.7, \,0) = (30.3, \,0, \,-21.2) \, \text{G}$. To avoid Majorana spin flips, we will apply another external field $B_z$ in the direction of the trapping wire to raise the minimum of the field to non-zero, to achieve the adiabatic limit. The external field $B_z = 0.5\,\text{G}$ together with $\textit{\textbf{B}}_{\text{bias}}$
form an IP trap with a non-zero minimum \cite{Abasalt:2019}.

\begin{figure}[htbp]
    \centering
    \includegraphics[scale=0.6]{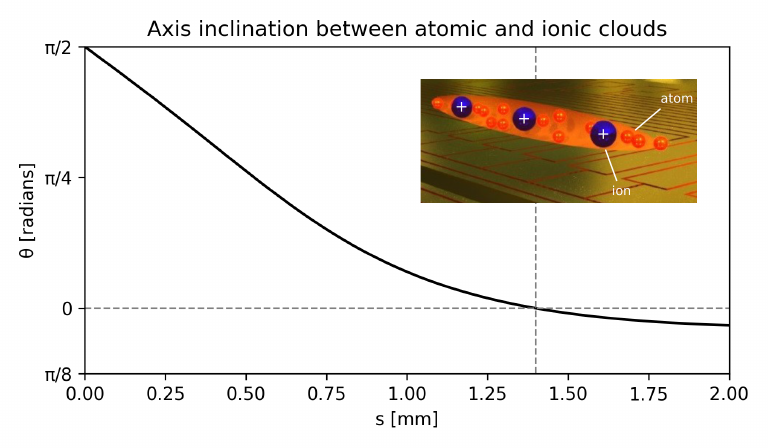}
    \caption{Angle $\theta$ between the z-axis of the ion trap and the axial direction of the atom trap potential versus the length of the z-shaped wire along z-direction at trapping height of 0.70\,mm. At $\theta$=0, ions represented by blue spheres, are immersed within the atom cloud.}
    \label{fig:z_wire_length}
\end{figure}

The axial orientation of the atom cloud confined within the IP trap may not align perfectly with the z-axis of the ion trap \cite{gerritsma2006topological}. In the design of the atom trap, we have taken into account a z-shaped wire length of s\,=\,1.4\,mm, ensuring significant overlap between the atom cloud and the ion cloud (Fig.~\ref{fig:z_wire_length}).

% The axial direction of the atom cloud trapped in the magnetic field of the z-shaped wire is not necessarily parallel to the z-axis of the ion trap. In the atom trap design, the length of the z-shaped wire is considered to be 1.4\,mm, so that the atom cloud overlaps with the ion cloud.

\subsection{Quadrupole magnetic field}
\begin{figure} [H]
	\centering
	\includegraphics[width=3in]{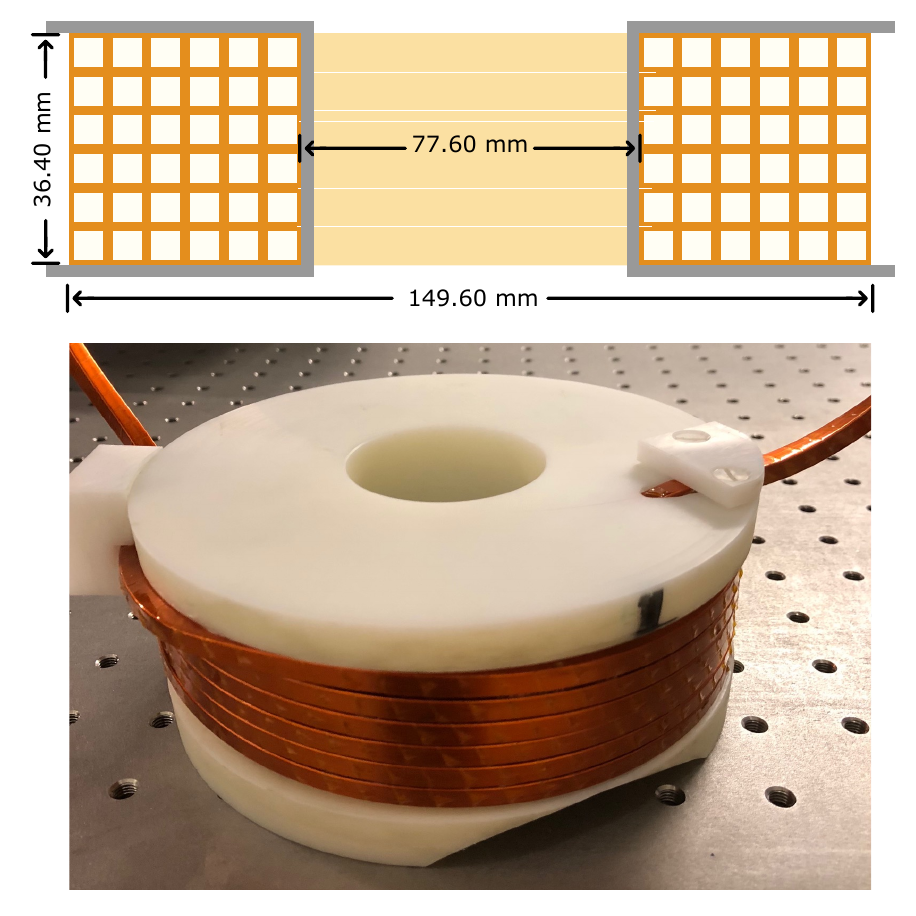}
	\caption{Sketch of one of the MOT coils. Each coil comprises 6$\times$6 turns made of a hollow-core copper wire with an outer and inner cross-sectional area of 6$\times$6\,mm$^{2}$ and 4$\times$4\,mm$^{2}$. The photograph below showcases a fabricated MOT coil.}
	\label{fig:MOTcoils}
	\end{figure}

The quadrupol magnetic field needed for the mMOT must be specifically engineered. A field that increases in strength as the distance from its center increases is necessary. Here, we discuss the design of a quadrupole magnetic field using two coils in an anti-Helmholtz configuration (Fig. \ref{fig:MOTcoils}). We constructed a set of MOT coils using hollow-core copper wire with a cross-sectional area of 6$\times$6\,mm$^{2}$ (inner cross-sectional area 4$\times$4\,mm$^{2}$). To ensure proper insulation between the wires, they were wrapped twice in Kapton tape.  Each coil comprises 36 turns and has an internal diameter of $77.60 \, \text{mm}$ and an external diameter of $149.60 \, \text{mm}$. During MOT operation, the maximum current ($I = 200 \, \text{A}$) provided by a power supply (\text{SM\,30-200}) flows through the wires of the coils. The current, denoted as $I$, produces a magnetic field gradient of $0.059 \times I \, \text{(G/cm)}$ in the axial direction and $0.029 \times I \, \text{(G/cm)}$ in the radial direction near the trap center.

% \begin{table}[ht]
%   \centering
%   \caption{Details of the individual MOT coils' production.}
%   \vspace{1em} % Adjust the space as needed
%   \begin{tabular}{@{} l @{\hspace{2em}} c @{\hspace{2em}} r @{}}
%     \hline
%     \textbf{Quantity} & \textbf{Value} & \textbf{Unit}\\
%     \hline
%     Number of winding & 36 & \\
%     Cross-section area & 20.0 & mm$^{2}$\\
%     Coil length & 11.99 & m \\
%     Coil mass & 2.14 & Kg \\
%     Water mass & 0.19 & Kg \\
%     Coil resistance & 10.0 & m$\Omega$ \\
%     Power dissipation & 419.0 & W \\
%     Voltage drop & 2.05 & V \\
%     Pressure drop & 0.5 & bar \\
%     Volumetric flow rate & 2.36 & l/min \\
%     Water mass flow rate & 2.35 & Kg/min \\
%     Fluid velocity & 2.46 & m/s \\
%     Reynolds number & 2575.4 & - \\
%     Temperature rise & 2.5 & $^{\circ} \mathrm{C}$ \\
%     \hline
%   \end{tabular}
%   \label{Tab:MOT_coils}
% \end{table}

\setlength{\tabcolsep}{10pt} % Adjust the value as needed
\renewcommand{\arraystretch}{1.2} % Default value: 1

\begin{table}[ht]
\centering
\caption{Details of the individual MOT coils' production.}
\vspace{0.1em} % Adjust the space as needed
\begin{tabular}{@{} l @{\hspace{3em}} c @{\hspace{3em}} r @{}}
\toprule
\textbf{Quantity} & \textbf{Value} & \textbf{Unit} \\ \midrule
    Number of winding & 36 & \\
    Cross-section area & 20.0 & mm$^{2}$\\
    Coil length & 11.99 & m \\
    Coil mass & 2.14 & Kg \\
    Water mass & 0.19 & Kg \\
    Coil resistance & 10.0 & m$\Omega$ \\
    Power dissipation & 419.0 & W \\
    Voltage drop & 2.05 & V \\
    Pressure drop & 0.5 & bar \\
    Volumetric flow rate & 2.36 & l/min \\
    Water mass flow rate & 2.35 & Kg/min \\
    Fluid velocity & 2.46 & m/s \\
    Reynolds number & 2575.4 & - \\
    Temperature rise & 2.5 & $^{\circ} \mathrm{C}$ \\

\bottomrule
\end{tabular}
\label{Tab:MOT_coils}
\end{table}

To estimate the thermal budget of the coils, we calculated the volumetric flow rate of the cooling water which is an incompressible liquid \cite{cornish1928flow}. The maximum pressure drop in our chiller is 4.5\,bar (MINORE 0-RB400). With a pressure loss of about 0.5\,bar, we estimate a temperature increase of about 2-4\,$^\circ$C, which is in close agreement with the measured temperature increase when the MOT coils were running continuously at $I$\,=\,200\,A. At this current, the energy dissipated in each coil is approximately 900\,W and the voltage loss 4.15\,V. The voltage loss increases to 7.0\,V when we engage the MOT switch (Fig.~\ref{fig:switch}).

\subsection{Fast switch for MOT coil}

\begin{figure}[ht]
	\centering
	\includegraphics[scale=0.35]{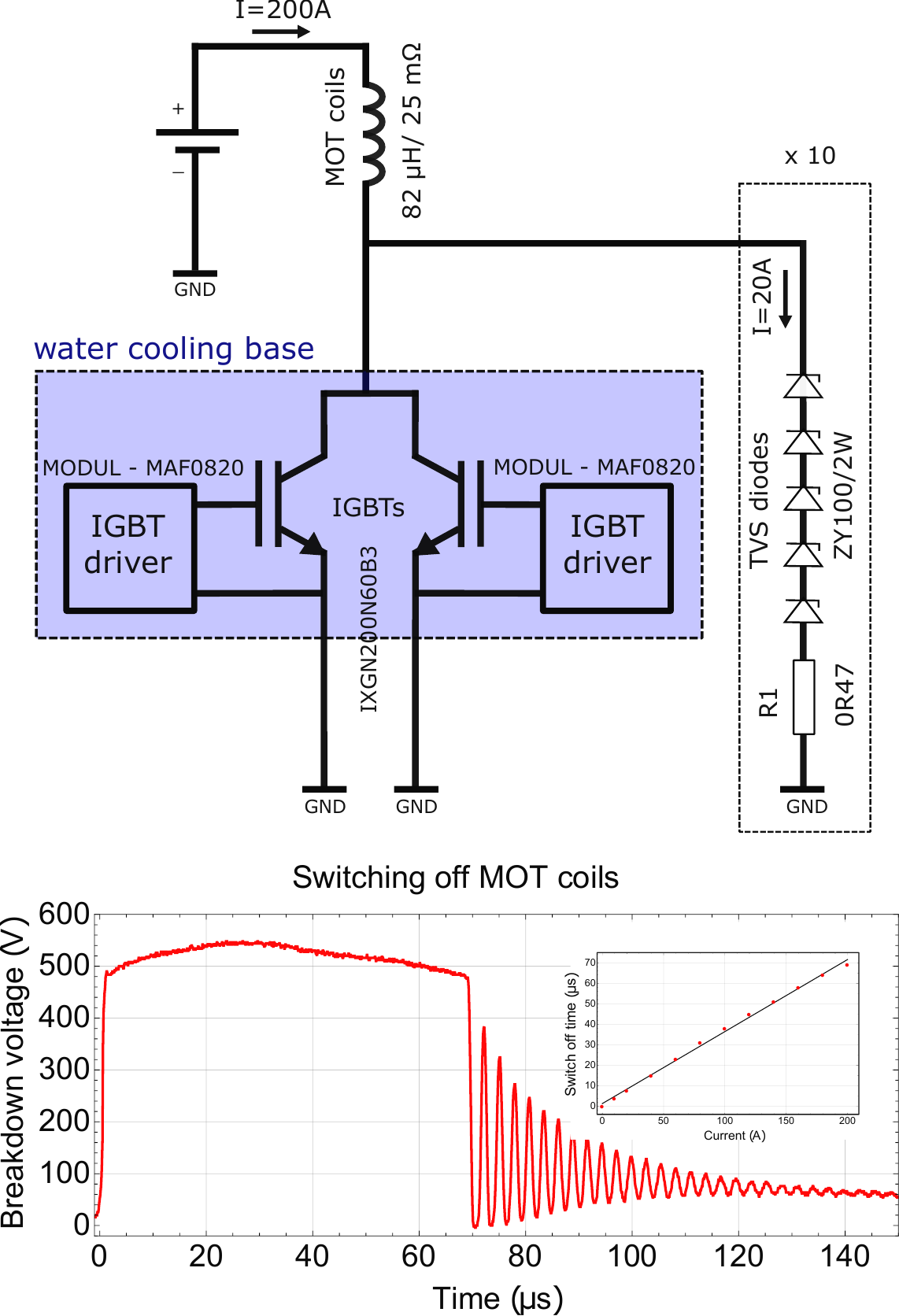}
	\caption{The circuit diagram for high-speed IGBT-driven current switching. The MOT coils are connected in series, resulting in a total electrical inductance of 82\,$\mu$H and a resistance of 25\,m$\Omega$. When the maximum current of 200\,A is switched off, an electromotive force is generated that triggers the transient-voltage-suppression diodes.}
	\label{fig:switch}
	\end{figure}

 Fast magnetic field switching is essential for maintaining the stability of the MOT, enhancing the laser cooling efficiency, and for performing various manipulation of trapped atoms. However, doing so caused eddy currents to form in the electrical conductive parts, leading to a slow decay of the magnetic field. To solve this issue, we implemented a Polyoxymethylene holder for the coils. To achieve fast switching, we developed a current driver using high-speed insulated gate bipolar transistors (IGBTs). We used a series connection of 10$\times$5 transient-voltage-suppressor diodes (TVS diodes) followed by a resistor (Fig.~\ref{fig:switch}). Each diode has a breakdown voltage of 100\,V, allowing the magnetic energy to dissipate to ground as soon as the reverse voltage reaches 500\,V. With this setup, we succeeded in achieving a switching-off time less than 100\,$\mu$s for 200\,A. We observed a linear relationship between the applied current in the magnetic trap coils and the switching-off time, with a rate of 0.45\,$\mu$s per 1\,A.

\section{ion trapping}\label{sec:ion-trapping}
% fsk: sect IV is on ion trapping. It would be great to have more material about the ion trap operation, linear crystals of different sort, isotopes, moving, mm compensation, ....more picture also

Trapping, cooling and imaging of $ \text{Yb}^+$ ions requires three lasers. The ionization of Yb atom is done via a two level scheme. A laser at 398.9\,nm excites the neutral Yb atom to the $^1$P$_1$ state. The second step is done with a laser at 369\,nm, which ionizes the Yb atom. The same laser beam is used to doppler cool and image the ion. This laser drives the $^2$S$_{1/2}\leftrightarrow ^2$P$_{1/2}$ dipole transition. In $0.5\%$ of the decays the ion falls into meta stable $^2$D$_{3/2}$ state. To bring the ion back to the cooling cycle a repump laser at 935\,nm is used \cite{Abasalt:2019}.

% It brings the ion to the short living $^3$D[3/2]$_{1/2}$ which decays back to the ground state.

% https://iopscience.iop.org/article/10.1088/1402-4896/ab635b/pdf

The blue lasers are overlapped and directed into a UV polarization-maintaining (PM) fiber, while the repump laser is coupled to an infrared (IR) PM fiber. Subsequently, all beams are combined using a dichroic mirror (Thorlabs M254C45), which reflects UV light and transmits IR light. A 200\,mm focal length achromatic lens is placed behind the dichroic mirror to focus the beams parallel to the ion chip surface in the ion trapping region, positioned around 100\,$\mu$m beneath the ion chip. The focal sizes of the UV beams at this location are 30\,$\mu$m, and the repump beam has a focal size of 100\,$\mu$m. The angle of incidence with respect to the ion-chip z-axis is $10^\circ$, enabling ion cooling in both radial and axial directions. Effective ion cooling is achieved by applying DC voltages to the DC electrodes E01-E20 (Fig.~\ref{fig:iontrap}) and rotating the principal axes of the trap~\cite{Abasalt:2019}.

\begin{figure}[h]
	\centering
	\includegraphics[scale=1]{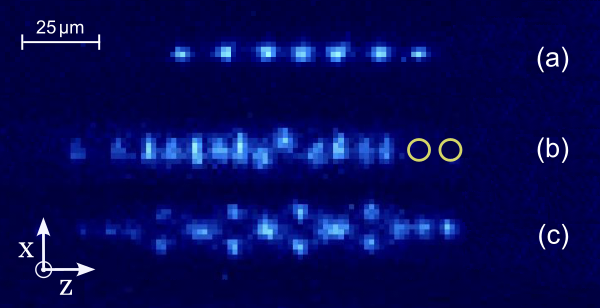}
	\caption{(a) Crystal of eight $^{174}$Yb$^{+}$ ions, (b) and (c) Zig-zag configuration of $^{174}$Yb$^{+}$ ions~\cite{Matthias}. By examining the positions of individual ions, we notice two dark Yb$^{+}$ ions enclosed in yellow circles. Each pixel in the image corresponds to $1.09\pm0.07$\,$\mu$m.}
	\label{fig:ion_crystals}
\end{figure}

After trapping ions, we used transverse confinement to arrange the trapped ions into a linear crystal. A gradual reduction in trap anisotropy triggers a structural phase transition, transforming the arrangement from linear to a zigzag configuration (Fig.~\ref{fig:ion_crystals}).

The lifetime of a Doppler cooled single ions are about 2\,min. Excessive micromotion amplitude can be addressed through various methods~\cite{narayanan2011electric, berkeland1998minimization, tanaka2012micromotion, allcock2012heating}. When a laser is red-detuned near the natural linewidth of the atom, the ion fluorescence rate increases with decreasing micromotion amplitude, particularly for low magnitudes of micromotion. After the 2D compensation, we measured a residual electric stray field uncertainty of \((\Delta \epsilon_{\text{x}}, \, \Delta \epsilon_{\text{z}}) = (0.76, \,0.45) \, \text{Vm}^{-1}\) at a trap frequency of \((\omega_x, \omega_z)/2 \pi =  (426, \,260) \, \text{kHz}\). This micromotion compensation is good enough for the initial studies of atom-ion interaction~\cite{cetina2012micromotion}.

\section{Atoms trapped in \MakeLowercase{m}MOT}

% fsk: sect V is on MOT trapping, here you can take a lot from existing text. The cavity and the Rb laser setup is in my view to long.

\begin{figure}[H]
    \centering
    \includegraphics[scale=1.6]{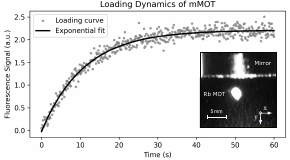}
    \caption{The dynamics of $^{87}$Rb atoms confined in a mMOT with a 13\,G/cm axial magnetic gradient. The solid line is the exponential fit to the measurements taken within a UHV chamber at a pressure of $1.3 \times 10^{-9}\, \mathrm{mbar}$. The bright spot in the center is the $^{87}$Rb MOT, and above this, the side of the gold-coated mirror is visible. The cooling laser beam intensity and detuning are \(6.5 \, \text{mW/cm}^2\) and \(-12 \, \text{MHz}\) respectively used in these experiments.}

    \label{fig:mMOT_mirror}
\end{figure}

Trapping atoms beneath the atom-ion chip is particularly challenging because the reflection of beams off the traps leads to changes in light polarization, alterations in beam quality, and variations in intensity.  Initially we tested the mMOT in a test setup with just a gold mirror (Thorlabs PF20-03-M03) instead of original trap. We aligned all the laser beams with the zero of the quadrupole magnetic field and trapped a mMOT directly from the background vapor of the atoms. Under our typical experimental conditions, 13\,G/cm axial magnetic field gradient and laser detuning of \(\Delta \approx -\Gamma/2\), we capture atomic cloud with an approximate radius of $r=(1.8\,\pm\,0.20)$\,mm that contains roughly \(N = (2.3 \pm 0.3) \times 10^8\) atoms of $^{87}$Rb. These values are determined from fluorescence signals (Fig.~\ref{fig:mMOT_mirror}). 

The fluorescence intensity emitted from trapped atoms in the MOT is proportional to the number of atoms. The dynamics of atoms loading into a MOT can be described by \cite{bjorkholm1988collision, prentiss1988atomic}:

\begin{equation}
\frac{dN(t)}{dt} = R - \gamma N(t) - \beta \bar{n} N(t).
\label{eq:atom_load}
\end{equation}

Here, \(N\) represents the atom count, \(R\) signifies the loading rate of atoms, \(\gamma\) denotes the loss rate resulting from spontaneous emission, and \(\beta\) accounts for the loss rate due to collisions with the background gas. The density of the background gas is represented as \(\bar{n}\). In the constant background gas density limit, Eq.~\ref{eq:atom_load} simplifies to:

\begin{equation}
N(t) = \frac{R}{(\gamma + \beta \bar{n})}(1-e^{-(\gamma + \beta \bar{n}) t}) = \frac{R}{\zeta}(1-e^{-\zeta t}).
\label{eq:mot_load_simple}
\end{equation}

We simplified Eq.~\ref{eq:mot_load_simple} by substituting $\gamma + \beta \bar{n}=\zeta$. By analyzing the loading curves of a mMOT as shown in Fig.~\ref{fig:mMOT_mirror}, one can determine the values of \(R\) and \(\zeta\). In our case, we typically obtain a ratio of \(R/\zeta  = (2.0 \pm 0.3) \times 10^8\) and a value of \( \zeta = 0.09 \pm 0.01 \, \text{s}^{-1} \).

\subsection{Big u-shaped wire trap}
Subsequently, we positioned the atom-ion trap inside the UHV chamber. We observed a 15\% reduction in the intensity of beams reflected from the ion trap surface due to microfabriacted structures on the chip. This imbalance in optical pressures results in a slight shift of the atom trapping location away from the zero point of the quadrupole magnetic field. However, the beam diameter (5-10\,mm) is sufficient to offset this effect. Following the recapturing of the mMOT beneath the ion-chip, we turned on the big u-wire potential by applying 15\,A (Fig.~\ref{fig:z_u_wires}).

\begin{figure}[htbp]
    \centering
    \includegraphics[scale=0.4]{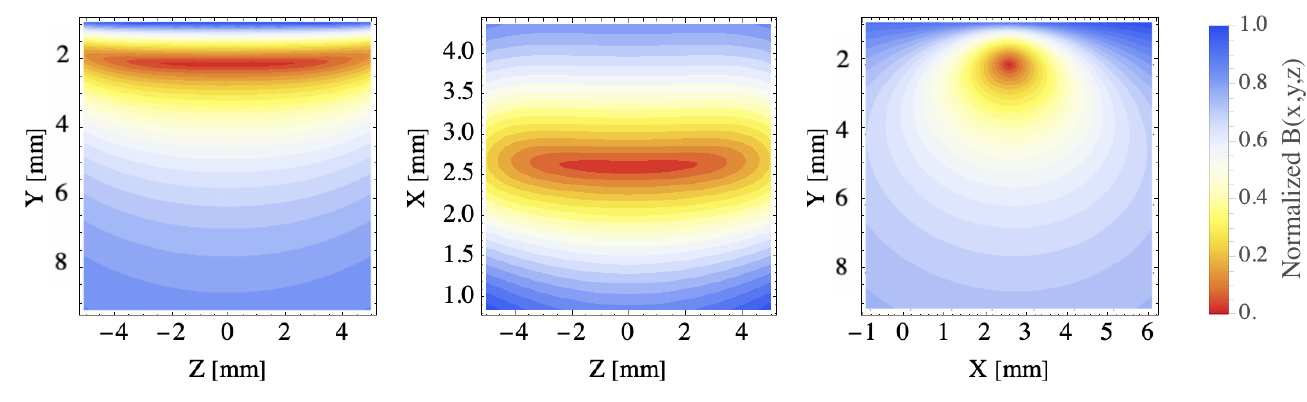}
    \caption{The zero-minimum magnetic field potential at (2.6, 2, 0)\,mm is generated by the big u-shaped wire carrying a current of I = 15\,A, along with bias fields of \( \textit{\textbf{B}}_{\text{bias}} = (10.3, 2.2, 0) \, \text{G} \).}
    \label{fig:big_u_potential}
\end{figure}

\begin{figure}[htbp]
    \centering
    \includegraphics[scale=1.1]{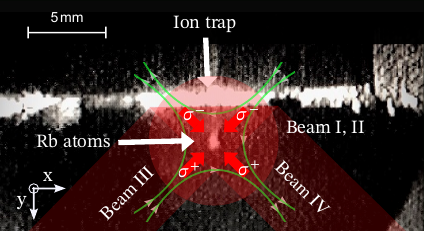}
    \caption{Image of the $^{87}$Rb mMOT under the ion trap chip. Beams I and II (directed toward/away from the page), and III are all RHC polarized, while beam IV is LHC polarized. The green lines depict the quadrupole magnetic field lines of the trapping potential created by big u-wire. The ion trap, with a width of $w=5$\,mm, is also visible due to light reflections.}
    \label{fig:mMOT}
\end{figure}

 In order to establish a field minimum at the coordinates $(x = 2.6,\, y = 2,\, z = 0)$\,mm, we applied bias fields $\textit{\textbf{B}}_{\text{bias}} = -\textit{\textbf{B}}_{\text{u-wire}}(2.6, 2, 0) = (10.3, 2.2, 0)\,\text{G}$ (Fig.~\ref{fig:big_u_potential}). This results in a magnetic field gradient of $\nabla\textit{\textbf{B}} = (4.8,\,-4.7,\, -0.1)\,\text{G/mm}$. To increase the intensities of the MOT beams, we decreased the beam diameter from 10\,mm to 5\,mm and trapped a relatively small atomic cloud positioned 2\,mm below the ion trap chip (Fig.~\ref{fig:mMOT}). At this position, the trapped atoms and ions are not yet spatially overlapped. The subsequent step involves optimizing the transfer of atoms from this location to the small u-shaped and ultimately z-shaped potentials.

A primary objective of this experiment is to investigate the interactions of atoms with ions through induced dipole moments. To this end, it is crucial to establish a characteristic range of the atom-ion interaction, which is defined by the length scale \(R^{*}=\sqrt{2C_{4}/\hbar^{2}}\). Here, \(R^{}\), the characteristic range, is determined by the constant \(C_{4}\) related to induced dipole moments and the reduced Planck's constant \(\hbar\).

For these interactions, the ionic wave packet length \(l_{a}=\sqrt{\hbar/m_{a}\omega_{a}}\) must be proportional with \(R^{}\), where \(l_{a}\) represents the ionic wave packet length, \(m_{a}\) the mass of the ions, and \(\omega_{a}\) the trap frequency for the ions. The specific mixture of \(^{174}\)Yb\(^{+}\) and \(^{87}\)Rb atoms used in this experiment results in \(R^{*}=306\,\text{nm}\), requiring a trap frequency of \(1.23\,\text{kHz}\). The corresponding single atom wave packet sizes are \(l_{a}=(315, 315, 1175)\,\text{nm}\). In contrast to traditional hybrid trap designs \cite{tomza2019cold}, this particular trap design offers improved trapping stability and a streamlined infrastructure, making it an ideal foundation for further advancements.

Having successfully integrated cooling and trapping techniques for atoms and ions on a single microfabricated chip, the subsequent phase involves confining the atoms in a MT to study atom-ion interactions. At low energy levels, two-body collisions between a hetero-nuclear ion \(\text{A}^+\) and an atom B can result in various processes. These include elastic collisions where the internal states of both the atom and ion remain unchanged \cite{cote2000classical}. Furthermore, spin exchange within the ion-atom pair is a potential occurrence \cite{selin2003single}. Additionally, in close proximity, an electron from atom B may transition to \(\text{A}^+\), initiating a charge transfer process \cite{zygelman1989radiative, zygelman1988radiative}.

\section{conclusion}
% fsk: sect VI is the text of the conclusion what is now in VI.

The hybrid atom-ion system we have introduced facilitates the exploration of Josephson physics. Using an atom-ion setup allows for the investigation of tunneling phenomena in a double well potential. The control over tunneling dynamics is achieved through manipulation of the internal state of the ion \cite{gerritsma2012bosonic}.  In scenarios where a crystal of ions is coupled to an atomic cloud, the ions organize into lattices, imparting a band structure onto the atoms. This coupling enables the exploration of various quantum phenomena, such as polaron physics and Peierls transitions \cite{bissbort2013emulating}. Furthermore, the detection of a magnetic field gradient induced by the magnetic dipole of a single atom becomes feasible with a single ion positioned nearby \cite{schmidt2012entangled}.

\section*{ACKNOWLEDGMENTS}
We acknowledge the scientific contributions of Rene Gerritsma, Matthias M{\"u}ller and Jannis Joger into the project. Our thanks go to Hartmut Häffner and his team for providing the trap chip. We express special thanks to the SFB/TR49 for their generous financial support.

\section*{AUTHOR DECLARATIONS}
\textbf{Conflict of Interest}

The authors have no conflicts to disclose.

\section*{DATA AVAILABILITY}
The data that support the findings of this study are available from the corresponding author upon reasonable request.

% \nocite{*}

\section*{REFERENCES}
\bibliography{aipsamp}% Produces the bibliography via BibTeX.

\end{document}